# Bridging Utility Maximization and Regret Minimization

Alessandro Chiesa    Silvio Micali    Zeyuan Allen Zhu

MIT

March 24, 2014

**Abstract**

We relate the strategy sets that a player ends up with after refining his own strategies according to two very different models of rationality: namely, utility maximization and regret minimization.

# 1 Introduction

Rational players have been modeled in two main ways.

- A utility-maximizing player $\mathcal{U}$ eliminates all his dominated strategies to compute his set of undominated ones, UD. Notice that $\mathcal{U}$ cannot further refine UD based on utility maximization. If UD consists of a single strategy $s$ (necessarily a dominant one), then $\mathcal{U}$ of course chooses $s$. *But, if UD contains multiple strategies, which one should $\mathcal{U}$ choose?*

- A regret-minimizing player $\mathcal{R}$ eliminates all his non regret-minimizing strategies so as to compute his set of regret-minimizing strategies, RM. He might even continue this process $k$ times, until he is satisfied or no further elimination is possible. Let us denote the final set of strategies he obtains this way by $\mathsf{RM}^k$. If $\mathsf{RM}^k$ consists of a single strategy $s$, $\mathcal{R}$ of course chooses $s$. *But, if $\mathsf{RM}^k$ contains multiple strategies, which one should $\mathcal{R}$ choose?*

In both cases, "a random strategy" or "the lexicographic first strategy" are certainly possible answers. But another answer is that, when he is 'no longer able to apply his favorite way of reasoning', even a die-hard utility maximizer $\mathcal{U}$ will resort to regret minimization to refine UD, and even a die-hard regret minimizer $\mathcal{R}$ will resort to utility maximization to refine $\mathsf{RM}^k$. In principle, the two final sets of strategies obtained by such different refinement procedures could be vastly different. Our next structural theorem, however, guarantees that they coincide.

Abusing notation a bit, consider UD and RM also to be "operators" acting on sets of strategies. In this case $\mathsf{UD}(\mathsf{UD}) = \mathsf{UD}$, while $\mathsf{RM}^2 \stackrel{\text{def}}{=} \mathsf{RM}(\mathsf{RM})$ may be a strict subset of RM. Then, we prove that the set of strategies obtained after applying, in arbitrary order, $k$ times the operator RM and at least once the operator UD coincides with $\mathsf{RM}^k \cap \mathsf{UD}$. For instance,

$$\mathsf{RM}(\mathsf{RM}(\mathsf{UD}(\mathsf{RM}(\mathsf{RM}(\mathsf{UD}))))) = \mathsf{RM}^4(\mathsf{UD}) = \mathsf{RM}^4 \cap \mathsf{UD}.$$

After recalling the relevant notions, we prove our theorem for pure strategies, and then point out its simple but interesting implications for mechanism design. Finally, we point out that our result extends to mixed strategies as well.

We recall that regret-minimizing strategies are also known as regret-minimax strategies. The suggestion of adopting regret-minimizing strategies traces back to Savage's reading [Sav51] of the work of Wald [Wal49], and has been axiomatized by Milnor [Mil54]. The notion of regret has been treated differently in different settings. A unified axiomatic characterization of minimax regret has been recently given by Stoye [Sto11].

Many empirical studies compare utility maximizers and regret minimizers, see for instance Chorus, Arentze and Timmermans [CAT09], and Hensher, Greene and Chorus [HGC11]. Recently, Engelbrecht-Wiggans and Katok [EK07] and Filiz and Ozbay [FO07] provide experimental evidence for regret in first- and second-price auctions.

To the best of our knowledge, we are the first to study players who use regret for refining their sets of undominated strategies.

## 2 Basic Notions

To state and prove our result, we use the language of decision theory: namely, envisaging "a single player against Nature".[1]

Let $\mathcal{S}$ be a compact set of (pure) strategies of the player, and $T$ a compact set of states of Nature.[2] We denote by $U$ the (continuous) utility function of the player, where $U(s,t)$ is the utility under strategy $s \in \mathcal{S}$ when Nature's state is $t \in T$. Regret-minimizing strategies and undominated strategies are defined as follows:

- Given a *menu* $S \subseteq \mathcal{S}$ of strategies, the player's (maximum) regret for a strategy $s \in S$ in menu $S$, denoted by $R_S(s)$, is the maximum difference, taken over all possible Nature's states $t \in T$, between the utility the player gets by playing $s$, and that he could have gotten by "best responding" to $t$; formally,

$$R_S(s) \stackrel{\text{def}}{=} \max_{t \in T} \Big( \max_{s^* \in S} U(s^*, t) - U(s, t) \Big).$$

---

[1] Results for $n$-player (strategic or pre-Bayesian) games follow as corollaries. This is because the definitions of dominance and regret are universally quantified over other players' strategies, which can be treated as Nature's strategies.

[2] Both $\mathcal{S}$ and $T$ may be infinite, and $\mathcal{S}$ may be convex in order to allow arbitrary mixed strategies to be considered.

Therefore, the *regret-minimizing strategies* with respect to a menu $S \subseteq \mathcal{S}$, denoted by $\mathsf{RM}(S)$, is the set of strategies that minimize the regret:

$$\mathsf{RM}(S) \stackrel{\text{def}}{=} \arg\min_{s \in S} R_S(s).$$

- Given two strategies $s, s' \in \mathcal{S}$, by definition $s'$ *weakly dominates* $s$, denoted by $s' \succ s$, if

$$\forall t \in T,\ U(s', t) \geq U(s, t) \quad \text{and} \quad \exists t \in T,\ U(s', t) > U(s, t) \ .$$

Given a *menu* $S \subseteq \mathcal{S}$ of strategies, the player's *undominated strategies* consist of those that are not weakly dominated by any weakly undominated strategy.[3] Formally,

$$\begin{aligned}
\mathsf{UD}(S) &\stackrel{\text{def}}{=} S \setminus \{s \in S \ :\ \exists s' \in S \text{ s.t. } (s' \succ s) \wedge (\nexists s'' \in S, s'' \succ s')\} \\
&= \quad \{s \in S \ :\ \nexists s' \in S \text{ s.t. } (s' \succ s) \wedge (\nexists s'' \in S, s'' \succ s')\}.
\end{aligned}$$

We now state two simple facts which easily follow from the above definitions:

**Fact 2.1.** *For any menu $\tilde{S} \subseteq \mathcal{S}$,*

(a) *if $s \prec s'$ for some $s, s' \in \tilde{S}$, then $R_{\tilde{S}}(s) \geq R_{\tilde{S}}(s')$, and*

(b) *the regret values of a strategy with respect to $\tilde{S}$ and $\mathsf{UD}(\tilde{S})$ are the same, namely:*[4]

$$R_{\tilde{S}}(s) = \max_{t \in T} \left( \max_{s^* \in \tilde{S}} U(s^*, t) - U(s, t) \right) = \max_{t \in T} \left( \max_{s^* \in \mathsf{UD}(\tilde{S})} U(s^*, t) - U(s, t) \right) = R_{\mathsf{UD}(\tilde{S})}(s) \ .$$

Note that regret minimization is mostly studied when a player has beliefs about his opponents. In particular, the notions from Hyafil and Boutilier [HB04] and Renou and Schlag [RS10] coincide with ours when the players do not form beliefs about their opponents —or, in our language, Nature.

---

[3]In general, weakly undominated strategies do not coincide with undominated ones. As argued by Jackson [Jac92], it may happen that every pure strategy is weakly dominated by another one in an infinite chain, and in such a case all strategies are undominated but weakly dominated. However, in many cases of interest (e.g., when the set of pure strategies is finite, or when the mechanism is *bounded*), weakly undominated strategies coincide with undominated ones.

[4]The equality in the middle is since any strategy $s^* \in \tilde{S} \setminus \mathsf{UD}(\tilde{S})$ must be weakly dominated by some $s^{**} \in \tilde{S}$, giving at least as good utilities as $s^*$ for *any* $t \in T$. Therefore, such choices of $s^{**}$ can be ignored in the inner max.

# 3 Result

Established our language, we prove our theorem as a corollary of the following lemma.

**Lemma 1.** *For any menu $S \subseteq \mathcal{S}$, $\mathsf{UD}(\mathsf{RM}(S)) = \mathsf{RM}(\mathsf{UD}(S)) = \mathsf{RM}(S) \cap \mathsf{UD}(S)$.*

*Proof.* We divide the proof into six steps:

1. $\mathsf{RM}(\mathsf{UD}(S)) \subseteq \mathsf{RM}(S)$.

    For any $s \in \mathsf{RM}(\mathsf{UD}(S))$, we show that $s \in \mathsf{RM}(S)$ by proving that $s$ has minimum regret among all strategies in $S$. Indeed:

    - For any other strategy $s' \in \mathsf{UD}(S)$, it holds that $R_{\mathsf{UD}(S)}(s) \leq R_{\mathsf{UD}(S)}(s')$. By Fact 2.1b, we deduce that $R_S(s) \leq R_S(s')$.

    - For any other strategy $s' \in S \setminus \mathsf{UD}(S)$, it holds that $s' \prec s''$ for some $s'' \in \mathsf{UD}(S)$ and $R_S(s) \leq R_S(s'')$. By Fact 2.1a, we deduce that $R_S(s) \leq R_S(s'') \leq R_S(s')$.

2. $\mathsf{RM}(\mathsf{UD}(S)) \subseteq \mathsf{UD}(\mathsf{RM}(S))$.

    Given that $\mathsf{RM}(\mathsf{UD}(S)) \subseteq \mathsf{RM}(S)$ (proved above), if there is some $s \in \mathsf{RM}(\mathsf{UD}(S))$ with $s \notin \mathsf{UD}(\mathsf{RM}(S))$, then $s$ must be weakly dominated by some other strategy $s' \in \mathsf{RM}(S)$, namely $s \prec s'$, but $s'$ cannot be weakly dominated by any other strategy in $\mathsf{RM}(S)$, by definition of $\mathsf{UD}$.

    Now we show that $s'$ cannot be weakly dominated by any strategy in $S$ as well. Suppose not, that is $s' \prec s''$ where $s'' \in S$. Then $s'' \notin \mathsf{RM}(S)$ as we have just argued. However, using Fact 2.1a we have $R_S(s') \geq R_S(s'')$, implying that $s'' \in \mathsf{RM}(S)$ since $s' \in \mathsf{RM}(S)$, giving a contradiction to $s'' \notin \mathsf{RM}(S)$.

    In sum, we showed that $s$ is weakly dominated by $s' \in S$, and in addition $s'$ cannot be weakly dominated by any strategy in $S$, contradicting the fact that $s \in \mathsf{UD}(S)$.

3. $\mathsf{UD}(\mathsf{RM}(S)) \subseteq \mathsf{UD}(S)$.

    Suppose not, that is, there exists some $s \in \mathsf{UD}(\mathsf{RM}(S))$ that is not in $\mathsf{UD}(S)$. By the definition of $\mathsf{UD}(S)$, the strategy $s$ must be weakly dominated by some $s' \in S$,

and in addition $s'$ cannot be weakly dominated by any other strategy in $S$. There are two cases here.

- The first case is when $s' \in \mathsf{RM}(S)$. This case is impossible because $s \in \mathsf{UD}(\mathsf{RM}(S))$ implies that if $s$ is weakly dominated by $s' \in \mathsf{RM}(S)$, then $s'$ must also be weakly dominated, contradicting the fact that $s'$ cannot be weakly dominated by any strategy in $S$.

- The second case is when $s' \notin \mathsf{RM}(S)$. Since $s \prec s'$, by Fact 2.1a we have $R_S(s) \geq R_S(s')$. However, because $s \in \mathsf{UD}(\mathsf{RM}(S))$ implies that $s \in \mathsf{RM}(S)$, it must hold that $s'$ is a regret minimizer with respect to $S$, contradicting the fact that $s' \notin \mathsf{RM}(S)$.

4. $\mathsf{UD}(\mathsf{RM}(S)) \subseteq \mathsf{RM}(\mathsf{UD}(S))$.

Given that $\mathsf{UD}(\mathsf{RM}(S)) \subseteq \mathsf{UD}(S)$ (proved above), consider any strategy $s \in \mathsf{UD}(\mathsf{RM}(S))$, and suppose that $s \notin \mathsf{RM}(\mathsf{UD}(S))$. Then there exists some $s' \in \mathsf{UD}(S)$ satisfying $R_{\mathsf{UD}(S)}(s) > R_{\mathsf{UD}(S)}(s')$. This implies, through Fact 2.1b, that $R_S(s) > R_S(s')$, contradicting the fact that $s \in \mathsf{RM}(S)$.

5. $\mathsf{RM}(\mathsf{UD}(S)) \subseteq \mathsf{RM}(S) \cap \mathsf{UD}(S)$.

Trivial given the previous steps: $\mathsf{RM}(\mathsf{UD}(S)) \subseteq \mathsf{UD}(S)$ and $\mathsf{RM}(\mathsf{UD}(S)) = \mathsf{UD}(\mathsf{RM}(S)) \subseteq \mathsf{RM}(S)$.

6. $\mathsf{RM}(S) \cap \mathsf{UD}(S) \subseteq \mathsf{RM}(\mathsf{UD}(S))$.

Take any strategy $s \in \mathsf{RM}(S) \cap \mathsf{UD}(S)$, and suppose that $s \notin \mathsf{RM}(\mathsf{UD}(S))$. Then there exists some $s' \in \mathsf{UD}(S)$ satisfying $R_{\mathsf{UD}(S)}(s) > R_{\mathsf{UD}(S)}(s')$. This implies, through Fact 2.1b, that $R_S(s) > R_S(s')$, contradicting the fact that $s \in \mathsf{RM}(S)$. □

It is not hard to see that Lemma 1 implies our theorem. That is,

**Theorem 1.** *From any menu $S \subseteq \mathcal{S}$, the set of strategies obtained by applying,* in arbitrary order, *$i$ times the operator* $\mathsf{RM}$ *and at least once the operator* $\mathsf{UD}$*, is:*

$$\mathsf{RM}^i(S) \cap \mathsf{UD}(S) \ .$$

# 4 Implications for Mechanism Design

Mechanism design enables a social planner to generate a desirable outcome by leveraging the rationality (and the beliefs) of the players. Most works in mechanism designs assume the players to be utility maximizers. In particular, implementation in undominated strategies traces back to Jackson [Jac92]. However, mechanism design also considers regret minimizers. In particular, Linhart and Radner [LR89] study regret-minimizing strategies in a sealed-bid mechanism for bilateral bargaining under complete information. Engelbrecht-Wiggans [Eng89] and Selten [Sel89] analyze first- and second-price sealed-bid auctions by incorporating regret for the bidders. Halpern and Pass [HP12] propose the solution concept of iterated regret minimization using beliefs, and argue that it actually is the only one capable of explaining the actual behavior of the players in some settings.

If a mechanism ensures that each player has a unique undominated strategy, then that strategy is also dominant, and thus the only regret-minimizing one. However, it is not always possible to design such mechanisms. The designer of a new mechanism $M$ may never be sure that $M$ will be played solely by utility-maximizing players, nor that it will be played solely by regret-minimizing players. In principle, if he designs $M$ so that it implements a social choice correspondence $f$ in undominated strategies, then $M$ might produce a non desired outcome when one of the players is a regret minimizer, and viceversa.

We wish to quickly point out that Theorem 1 has an immediate but reassuring consequence for mechanism design.

> *Assume that a mechanism $M$ implements a social choice correspondence $f$ whenever each player chooses a strategy is a strategy subset that coincides either with* RM *or with* UD*. Then $M$ is automatically guaranteed to implement $f$ whenever each player chooses a strategy in his set* RM(UD)*.*

For instance, a mechanism implementing $f$ for regret minimizers continues to implement $f$ when the players are utility maximizers who resort to regret only for further refining, if needed, their sets of undominated strategies.

# 5  Pure vs. Mixed Strategies

So far we have been ambiguous, when discussing undominated strategies and regret-minimizing ones, about whether or not the players consider only pure strategies or also mixed ones. When only pure strategies are allowed, a utility maximizer compares only between his pure strategies for the notion of dominance and plays a pure undominated one, while a regret minimizer picks a pure strategy that minimizes regret among his pure strategies.

Our theorem and lemma are stated for pure strategies.

When mixed strategies are allowed, the definitions of UD and RM need more careful attention. It is easy to see that, when considering mixed strategies for regret minimizers, the only change needed is to allow such a minimizer to choose a mixed strategy that minimizes his expected regret among all his mixed ones (see e.g., [HB04, HP12]). Note that, it is easy to construct examples in which a mixed strategy yields strictly smaller regret than any pure strategy.

It is important to realize, however, that if we allow regret minimizers to consider mixed strategies, we *should* also allow utility maximizers to consider mixed strategies. For instance, our structural lemma (Lemma 1) would have difficulty to equate a set of pure strategies and a set of mixed ones. A utility maximizer may consider mixed strategies when determining that a strategy $s$ is weakly dominated by another strategy $s'$. The two interesting cases to consider are (1) $s$ is pure and $s'$ is mixed; and (2) both $s$ and $s'$ are mixed. Traditionally, most attention has been devoted to the first case, but the second has been studied too (see for instance [CS05, RS10]). Clearly, UD can be defined in both cases, and yields a more "refined" set of strategies in the second case.[5] It is actually under this more refined case that our structural lemma holds. In a sense, we have nothing to lose and something to gain by adopting a more flexible definition, after all the right notions are those yielding the right theorems.

---

[5] Let $\mathsf{UD}^{\mathsf{pure}}$ be the set of (pure) undominated strategies in the first case, and UD be the set of (possibly mixed) undominated strategies in the second case. Then, UD is a more "refined" notion of undominated strategies than $\mathsf{UD}^{\mathsf{pure}}$ because $\mathsf{UD}^{\mathsf{pure}} \subseteq \mathsf{UD} \subseteq \Delta(\mathsf{UD}^{\mathsf{pure}})$, i.e., $\mathsf{UD}^{\mathsf{pure}}$ coincides with the support of UD. For this reason, there is no difference in choosing between the two notions in most of the literature (see [CS05, footnote 2]).